\documentclass[preprint,number]{elsarticle}
\usepackage{graphicx}
\begin{document}

\begin{frontmatter}
\title{Data preservation at the Fermilab Tevatron}
\author[pado]{S. Amerio}
\author[fnal]{S. Behari}
\author[fnal]{J. Boyd}
\author[wash]{M. Brochmann}
\author[fnal]{R. Culbertson}
\author[fnal]{M. Diesburg}
\author[fnal]{J. Freeman}
\author[fnal]{L. Garren}
\author[fnal]{H. Greenlee}
\author[fnal]{K. Herner}
\author[fnal]{R. Illingworth}
\author[fnal]{B. Jayatilaka\corref{cor1}}
\author[fnal]{A. Jonckheere}
\author[fnal]{Q. Li}
\author[fnal]{S. Naymola}
\author[fnal]{G. Oleynik}
\author[fnal,roc]{W. Sakumoto}
\author[zona]{E. Varnes} 
\author[fnal]{C. Vellidis}
\author[wash]{G. Watts}
\author[fnal]{S. White}
\address[pado]{{\em Istituto Nazionale di Fisica Nucleare, Sezione di Padova-Trento and University of Padova, I-35131 Padova, Italy}}
\address[fnal]{{\em Fermi National Accelerator Laboratory, Batavia, Illinois, 60510, USA} }
\address[wash]{{\em University of Washington, Seattle, Washington, 98195, USA} }
\address[roc]{{\em University of Rochester, Rochester, New York 14627, USA}}
\address[zona]{{\em University of Arizona, Tuscon, Arizona, 85721, USA} }
\cortext[cor1]{Corresponding author. E-mail address: boj@fnal.gov}

\begin{abstract}
The Fermilab Tevatron collider's data-taking run ended in September 2011, yielding a dataset with rich scientific potential. The CDF and D0 experiments each have approximately 9 PB of collider and simulated data stored on tape. A large computing infrastructure consisting of tape storage, disk cache, and distributed grid computing for physics analysis with the Tevatron data is present at Fermilab. The Fermilab Run II data preservation project intends to keep this analysis capability sustained through the year 2020 and beyond. To achieve this goal, we have implemented a system that utilizes virtualization, automated validation, and migration to new standards in both software and data storage technology and leverages resources available from currently-running experiments at Fermilab. These efforts have also provided useful lessons in ensuring long-term data access for numerous experiments, and enable high-quality scientific output for years to come. 
\end{abstract}

\end{frontmatter}

\section{Introduction}

The Tevatron was a proton-antiproton collider located at Fermi National Accelerator Laboratory (Fermilab). Run II of the Tevatron, occurring from 2001 to 2011 and having collisions with a center-of-mass energy of 1.96 TeV, saw the CDF and D0 collaborations \cite{CDFdet1,CDFdet2,D0det} record datasets corresponding to an integrated luminosity of approximately 11 fb$^{-1}$ per experiment. These datasets helped make groundbreaking contributions to high energy physics including the most precise measurements of the $W$ boson and top quark masses, observation of electroweak production of top quarks, observation of $B_s$ oscillations, and first evidence of Higgs boson decay to fermions.  
The unique nature of the Tevatron's proton-antiproton collisions and large size of the datasets means that the CDF and D0 data will retain their scientific value for years to come, both as a vehicle to perform precision measurements as newer theoretical calculations appear, and to potentially validate any new discoveries at the LHC.

The Fermilab Run II Data Preservation Project (R2DP) aims to ensure that both experimental collaborations have the ability to perform complete physics analyses on their full datasets through at least the year 2020. To retain full analysis capability, the project must preserve not only the experimental data themselves, but also their software and computing environments. This requires ensuring that the data remain fully accessible in a cost-effective manner and that experimental software and computing environments are supported on modern hardware. Furthermore user jobs must be able to run at newer facilities when dedicated computing resources are no longer available and job submission and data movement to these new facilities must be accomplished within the familiar software environment with a minimal amount of effort on the part of the end-user. Documentation is also a critical component of R2DP and includes not only the existing web pages, databases, internal documents, but also requires writing clear, concise instructions detailing how users need to modify their usual habits to work in the R2DP computing infrastructure.

\section{Dataset preservation}

\subsection{Collision data}
\begin{figure}[ht]
\begin{center}
\includegraphics[width=1.0\columnwidth]{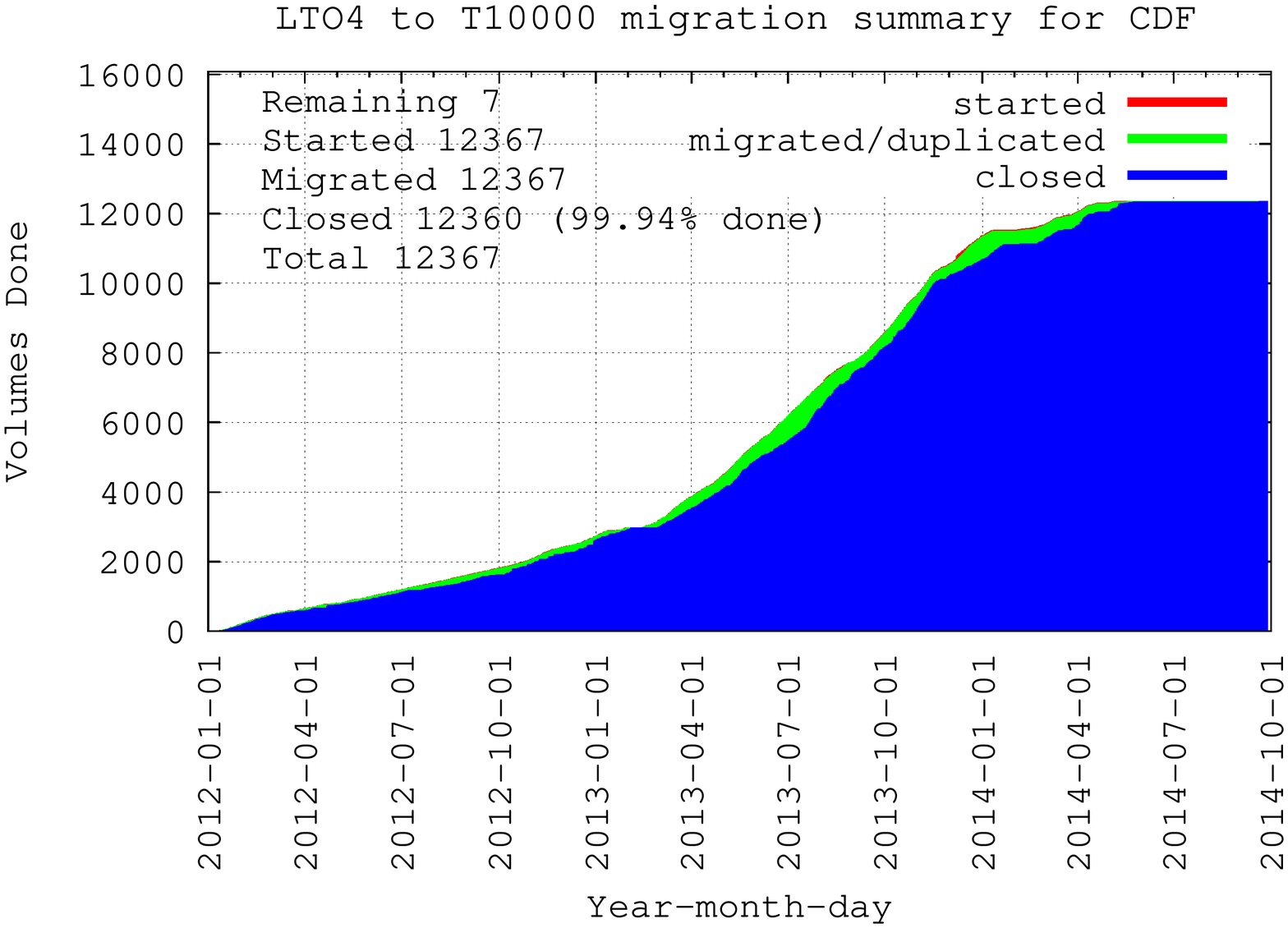}
\includegraphics[width=1.0\columnwidth]{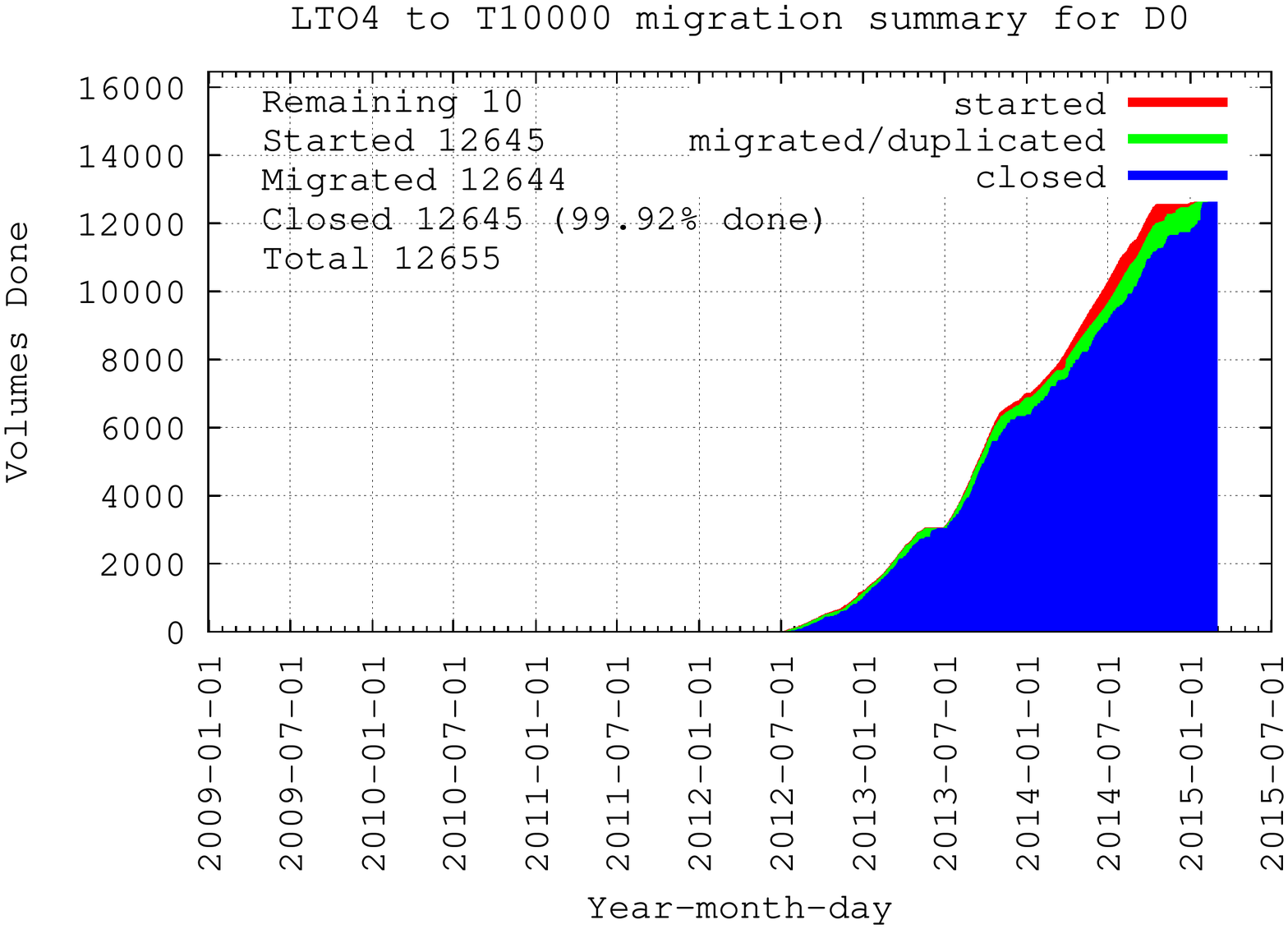}
\caption{Tape migration status from LT04 media to T10K media over the course of R2DP for CDF (top) and D0 (bottom). } 
\label{fig:tapemigration}
\end{center}
\end{figure}

At the time of the Tevatron shutdown, the data for both CDF and D0 were stored on LTO4 tapes \cite{lto}, which have a per-tape capacity of 800~GB. An analysis of then-available tape technologies concluded that T10K tapes \cite{t10k}, with a capacity of up to 5~TB per tape, would be the near-term choice for archival storage at Fermilab. While it was theoretically possible to leave the CDF and D0 data on LTO4 storage, a decision was made to migrate these data to T10K storage for two reasons. First, if the Tevatron data were accessed for a long period of time after data taking ended, the LTO4 storage may be an unsupportable configuration; as LTO4 tapes decline in usage industry-wide there may not be replacement storage easily available.  Second, as storage media and drives for older technology become scarce, their costs rise, potentially increasing the overall long-term cost of staying with LTO4 storage. Due to these concerns, the commitment was made to purchase T10K tapes and migrate all of the CDF and D0 data. It took approximately two years for the migration to be completed (Fig.~\ref{fig:tapemigration}) and the CDF and D0 data now share tape access resources with active Fermilab experiments.
Both CDF and D0 migrated all data stored on tape, including raw detector data, reconstructed detector data and derived datasets, and simulation. CDF has also made an additional copy of its raw detector data at the National Centre for Research and Development in Information Technology (CNAF) in Italy~\cite{cnaf}. Table~\ref{tab:datatypes} shows the amount of each type of data that CDF and D0 have stored over Run II. 
\footnote{While not within the scope of R2DP, a copy of raw CDF data from Run 1 of the Tevatron (1992-1996) is also being made at CNAF.}

\begin{table}
\begin{center}
\begin{tabular}{lcc}
Data type & CDF & D0 \\
\hline \hline
Raw collision & 1.8 & 2.5 \\
Reconstructed collision & 3.5 & 3.4 \\
Derived collision & 1.5 & 1.4 \\
Simulation & 2.0 & 1.5 \\
\hline
Total & 8.8 & 8.8 \\ 
\end{tabular}
\caption{Types of data stored on tape for CDF and D0 along with approximate size in PB. The numbers do not include the additional 4 PB of CDF raw data stored at CNAF.}
\label{tab:datatypes}
\end{center}
\end{table}

\subsection{Non-statistical data}
Throughout the Tevatron run, both CDF and D0 used Oracle database software for non-statistical data, such as detector calibrations. The ongoing cost of maintaining Oracle licenses, which are not used by most current Fermilab experiments for scientific use, presented a long-term challenge. The database schema was heavily interwoven into the analysis software.  As a result, converting to a more economical open source database solution such as PostgreSQL would incur a prohibitive investment in human resources. Thus, both experiments decided to retain the Oracle database systems throughout the life of the data preservation period, following an upgrade to the most recent version of Oracle at the time of the Tevatron shutdown.  Furthermore, as future upgrades to the Oracle database could potentially disrupt the existing schema, and thus the analysis software, a contingency plan was drawn up where the current version and schema could be frozen and run, in network isolation if necessary, in the future, even if support for that version had ceased. The CDF and D0 Oracle database schemas currently contain $1.89\times 10^{10}$ rows and continue to be accessed by physicists generating simulated data for ongoing analyses. A breakdown of the main types and size of data in the CDF and D0 databases appears in Table~\ref{tab:dbrows}. While the project decided to retain the Oracle databases, there was a risk that the physical hardware could fail before the end of the project lifetime. To mitigate that risk, the experiments moved the databases themselves to virtual machines. Most of the servers used for database hosting and caching at both experiments were transitioned to virtual hardware with no degradation in performance or uptime.

\begin{table}
\begin{center}
\begin{tabular}{c|c|c}
Database & Number of rows & Size on disk\\
 & (millions) & (GB) \\
\hline \hline
CDF SAM & 1\,015 & 201.6 \\ 
CDF Runs & 1\,336 & 195.6 \\
CDF subdetector calib. & 643 & 187.7 \\
CDF file catalog & 112 & 24.7 \\ 
D0 Luminosity & 5\,162 & 973.2 \\
D0 Runs & 85 & 21.8 \\
D0 SAM & 9\,178 & 2\,222 \\
D0 subdetector calib. & 901 & 88.5 \\ 
\end{tabular}
\caption{Summary of CDF and D0 databases. SAM stands for Sequential Access via Metadata (see Sec.~\ref{sec:data}). CDF and D0 internal database structures somewhat differ, leading to different sizes for the major database categories such as SAM and Runs.}
\label{tab:dbrows}
\end{center}
\end{table}

\section{Software and environment preservation}

Both CDF and D0 have complex software frameworks to carry out simulation, reconstruction, calibration, and analysis. Most of the core software was developed in the early- to mid-2000s on Scientific Linux running on 32-bit x86 architectures. During the operational period of the Tevatron, releases of the software framework were maintained on dedicated storage elements that were mounted on the experiments' respective dedicated computing clusters. As these dedicated resources are no longer maintained, CDF and D0 have migrated their software releases to CERN Virtual Machine File System (CVMFS) repositories~\cite{cvmfs}. As CVMFS has been widely adopted by current Fermilab experiments, this move allows for maintaining CDF and D0 software releases for the foreseeable future without a significant investment in dedicated resources. Furthermore, as the Fermilab-based CVMFS repositories are distributed to a variety of computing facilities away from Fermilab, this approach lends to CDF and D0 computing environments that exist on a variety of remote sites. 

At the time of the Tevatron shutdown, both CDF and D0 were running software releases that, while operational on Scientific Linux 5 (``SL5''), depended on compatibility libraries that were built in previous OS releases dating back to Scientific Linux 3. The two experiments chose different strategies to ensure the functionality of their software releases throughout the data preservation period.

\subsection{CDF software release preservation}
At CDF, stable software releases were available under two flavors: one was used in reconstruction of collision data and analysis and another for Monte Carlo generation and simulation. To ensure the long-term viability of CDF analysis capability, the CDF software team chose to prepare brand new ``legacy'' releases of both flavors. These legacy releases were stripped of any long-obsolete packages that were no longer used for any analysis and also shed any compatibility libraries built prior to SL5. Once validated and distributed, older releases of CDF software which still depended on compatibility libraries were removed from the CVMFS repository. This meant that usage of CDF code for analysis on any centrally available resources was guaranteed to be fully buildable and executable on SL5. Furthermore, it meant a relatively simple process for further ensuring the legacy release is buildable and executable exclusively on Scientific Linux 6 (``SL6''), the target OS for R2DP. The total size of the CDF code base
in the legacy releases is 326~GB, which includes compiled code and most external dependencies.

\subsection{D0 software release preservation}

After careful study, the D0 software team chose to stay with its software releases that were current at the time of the Tevatron shutdown, but updated common tools where possible, and also made sure that 32-bit compatibility system libraries are installed on worker nodes at Fermilab, where D0 plans to run jobs throughout the R2DP project lifetime. If D0 physicists should wish to run analysis jobs outside of Fermilab in future years, they will need to ensure that 32-bit versions of system libraries such as GLIBC are available at any future remote sites. Resources at Fermilab, however, are sufficient to meet the projected demand over the project lifetime. Required pre-SL6 compatibility libraries can also be added to the CVMFS repository if needed in the future. The total size of the D0 software repository in CVMFS, including code base, executables, and external product dependencies, is currently 227~GB.

\section{Job submission and data movement}

\subsection{Job submission}

\begin{figure}
\begin{center}
\includegraphics[width=1.0\columnwidth]{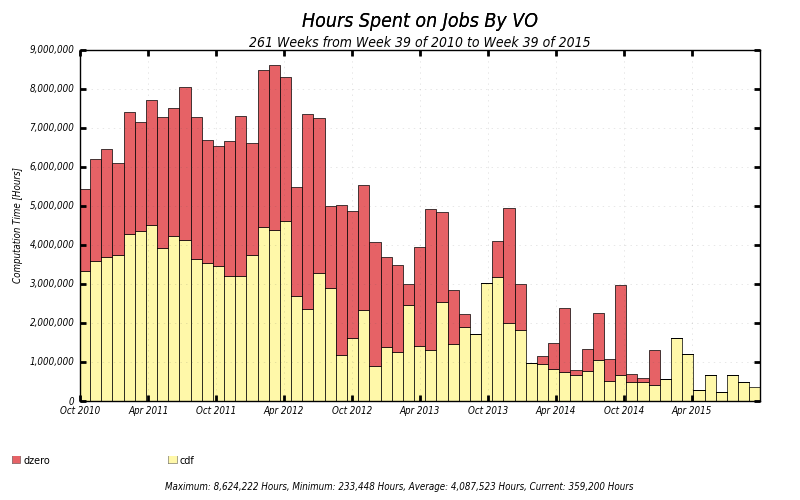}
\caption{Monthly utilization of computational resources (in core-hours) on the Open Science Grid by CDF (light yellow) and D0 (dark red) from October 2010 to October 2015. These resources include CDF and D0 dedicated resources prior to 2015 as well as GPGrid resources at Fermilab. Since early 2015 the bulk of D0 computing has run on its dedicated CAB cluster, which is not shown here from 2015 onward.}
\label{fig:gridjobs}
\end{center}
\end{figure}

During Run II, CDF and D0 both had large dedicated analysis farms (CDFGrid and CAB, respectively) of several thousand CPU cores each. Since the end of the run, these resources have been steadily diminishing as older nodes are retired and some newer ones are repurposed. While the computing needs of the experiments have declined over the years (Fig.~\ref{fig:gridjobs}), preserving full analysis capability requires that both experiments have access to opportunistic resources and a way to submit jobs to them. The Fermilab General Purpose Grid (``GPGrid''), used by numerous other experiments based at Fermilab, is a natural choice for the Tevatron experiments. Both CDF and D0 have worked with the Fermilab Scientific Computing Division to add the ability to run their jobs on GPGrid, by adopting the Fermilab Jobsub product~\cite{jobsub} used by other Fermilab experiments. Having users submit their analysis jobs via Jobsub solves the issue of long-term support, but introduces an additional complication for users who are unfamiliar with the new system, or who may return to do a Tevatron analysis many years from now and will not have time to learn an entirely new system. Thus, both CDF and D0 have implemented wrappers around the Jobsub tool that emulate job submission commands each experiment normally uses.

In the case of D0, users who wish to submit to GPGrid instead of CAB (which will be required once CAB is retired) can simply do so by adding an extra command line option. The D0 submission tools will then generate and issue the appropriate Jobsub commands without any direct user intervention. Users can switch to submitting jobs to GPGrid with a minimum amount of effort, and future analyzers will not need to spend time learning an entirely new system. We have successfully tested submission of all common job types (simulation, reconstruction, user analysis) to GPGrid using the modified D0 submission tools. 

In the case of CDF, the existing CDFGrid gateway was retired with all remaining hardware (worker nodes) absorbed into GPGrid. Analysis and Monte Carlo generation jobs are now submitted entirely using Jobsub and a CDF-specific wrapper in front. This also allows for CDF analysis jobs to go to remote sites on the Open Science Grid, specifically to sites that previously had to operate CDF-specific gateways. These sites can now support CDF computing either opportunistically or via dedicated quotas without the need to support a separate gateway. This transition was completed in early 2015 and CDF computing use did not diminish in 2015 as compared to 2014 despite moving to an environment with no dedicated computing nodes. CDF physicists consumed over 5 million CPU hours on GPGrid in the twelve months following the transition. 

\subsection{Data management and file delivery\label{sec:data}}

Both CDF and D0 use the Sequential Access to Metadata (SAM) service~\cite{sam} for data handling. Older versions of SAM used Oracle backends with a CORBA-based communication infrastructure, while more recent versions use a PostgreSQL-based backend, with communication over http. Throughout Run II CDF and D0 used CORBA-based versions of SAM, but can now also communicate with their existing Oracle databases using the new http interfaces as part of R2DP, eliminating the requirement of supporting the older CORBA-based communication interface through the life of the project. 
While CDF and D0 code bases had to be modified to interface with these new communication interfaces, these 
changes are transparent to the end user.

For D0, part of the SAM infrastructure included dedicated cache disks on the D0 cluster worker nodes that allowed for rapid staging of input files to jobs. As files were requested through SAM, they would be copied in from one of the cache disks if they were present. If they were not already on one of the cache disks SAM would fetch them from tape. At its peak this cache space totaled approximately 1~PB, but this cache space was only available to D0 and would have been too costly to maintain over the life of the data preservation project. D0 has therefore deployed a 100 TB dCache~\cite{dcache} instance for staging input files to worker nodes, as CDF and numerous other Fermilab experiments are already doing. The test results showed no degradation in performance relative to the dedicated SAM caches, and the D0 dCache instance has been in production for approximately two years.  

As CDF was already using dCache for tape-backed caching, once the necessary code changes were made to use newer versions of SAM, data access continued to be possible with no hardware infrastructure changes needed. 

\section{Documentation}
Preserving the experiments' institutional knowledge is a critically important part of the project. Here we define this knowledge to be internal documentation and notes, presentations in meetings, informational web pages and tutorials, meeting agendas, and mailing list archives. The largest step in this part of the project was transferring each experiment's internal documents to long-term repositories. Both CDF and D0 have partnered with INSPIRE~\cite{inspire} to transfer their internal notes to experiment-specific accounts on INSPIRE.  
For internal meeting agendas, D0 has moved to an Indico instance hosted by the Fermilab Scientific Computing Division, while CDF has virtualized their MySQL-based system. Fermilab will see to it that archives of each experiment's mailing lists are available through the life of the project, and Wiki/Twiki instances are being moved to static web pages to facilitate ease of movement to new servers if needed in the future. In addition to moving previous documentation to modern platforms, both experiments have written new documentation specifically detailing the infrastructure changes that the R2DP project has made, along with instructions for adapting
legacy workflows to the new systems.

\section{Summary}

The Run II Data Preservation Project aims to enable full analysis capability for the CDF and D0 experiments through at least the year 2020. Both experiments have modernized
their software environments and job submission procedures in order to be able to run jobs in current operating systems and to take advantage of non-dedicated computing resources. 
Wherever possible they have adopted elements of the computing infrastructure now in use by the majority of active Fermilab experiments. They have also made significant efforts at preserving
institutional knowledge by moving documentation to long-term archives. 
Materials costs for the project were dominated by media for migrating the data to T10K tapes, while the vast majority of the project's other costs came from salaries. The implementation phase of the project is complete and both experiments are actively using the R2DP infrastructure
for their current and future work.

\section*{Acknowledgments}
The authors thank the computing and software teams of both CDF and D0 as well as the staff of the Fermilab Computing Sector that made this project possible.
 Fermilab is operated by Fermi Research Alliance, LLC under Contract number DE-AC02-07CH11359 with the United States Department of Energy. The work was supported in part by 
 Data And Software Preservation for Open Science (DASPOS) (NSF-PHY-1247316).


\end{document}